\documentclass{aa}

\usepackage{graphicx}
\usepackage{txfonts}
\usepackage{epsfig}

\begin{document}

\title{High resolution optical spectroscopy of IRAS 09425$-$6040 (=GLMP 260)\thanks{Based on observations collected at the European
Southern Observatory (La Silla, Chile), on observations made with ISO, an ESA
project with instruments funded by ESA Member States (especially the PI
countries: France, Germany, the Netherlands and the United Kingdom) with the 
participation of ISAS and NASA, and on observations made with the NASA/ESA 
Hubble Space Telescope, obtained from the data Archive at the Space Telescope 
Science Institute, which is operated by the Association of Universities for
Research in Astronomy, Inc., under NASA contract NAS5-26555} }

\author{D. A. Garc\'\i a-Hern\'andez\inst{1}\and C. Abia\inst{2}\and A. 
Manchado\inst{3,4}\and P. Garc\'\i a-Lario\inst{1}}
\institute{
ISO Data Centre, European Space Astronomy Centre (ESAC), ESA. Villafranca del 
Castillo. P.O. Box - Apdo. 50727. E-28080 Madrid. Spain
\and Departamento de F\'\i sica Te\'orica y del Cosmos, Universidad de 
Granada, E-18071 Granada, Spain.
\and Instituto de Astrof\'\i sica de Canarias, E-38200 La Laguna, Spain.
\and Consejo Superior de Investigaciones Cient\'\i ficas, Spain.}

\date{Received February 11, 2006 / Accepted March 10, 2006}
\offprints{D. A. Garc\'\i a-Hern\'andez, e-mail: {\rm Anibal.Garcia@sciops.esa.int}}

\abstract{
We present high resolution optical spectroscopic observations of IRAS
09425$-$6040, a peculiar, extremely red, C-rich AGB star showing prominent
O-rich dust features in its ISO infrared spectrum attributed to crystalline
silicates. Our analysis shows that IRAS 09425$-$6040 is indeed a C-rich star
slightly enriched in lithium (\textit{log (Li/H) + 12} $\sim$ 0.7) 
with a low
$^{12}$C/$^{13}$C = 15$\pm$6 ratio. We also found some evidence that it may be
enriched in s-elements. Combining our results with other observational data
taken from the literature we conclude that the star is possibly an
intermediate-mass TP-AGB star (M $\gtrsim$ 3 M$_\odot$) close to the end of 
its AGB
evolution which may have only very recently experienced a radical change in its
chemistry, turning into a carbon-rich AGB star. \keywords{AGB and post-AGB stars - stars: individual: IRAS
09425$-$6040 (=GLMP 260) - stars: abundances} }

\authorrunning{Garc\'\i a-Hern\'andez et al.}
\titlerunning{IRAS 09425$-$6040}
\maketitle

\section{Introduction}
The evolution of low- and intermediate-mass stars (0.8 $\leq$ M $\leq$ 8
M$_\odot$) ends with a phase of strong mass loss on the Asymptotic Giant Branch
(AGB) phase. The chemical appearance (C-rich or O-rich) of these stars during
the AGB phase depends mainly on the progenitor mass and metallicity. Low-mass
AGB stars (M $\lesssim$ 2$-$3 M$_\odot$), initially O-rich, can switch to a 
C-rich chemistry (C/O$>$1 in the envelope) after a certain number of thermal
pulses and the subsequent dredge-up of C-rich material to the surface of the
star. Higher mass stars (M $\gtrsim$ 3$-$4 M$_\odot$), in contrast, remain
O-rich during their whole AGB evolution, due to the activation of the  ``Hot
Bottom Burning'' (HBB; e.g. Sackmann \& Boothroyd 1992; Mazzitelli, D'Antona \&
Ventura 1999) process, which prevents the formation of carbon, favouring the
production of nitrogen, instead.  HBB takes place when the temperature at the
base of the convective envelope is hot enough (T $\geq$ 2$\times$10$^{7}$ K)
that $^{12}$C can be converted into $^{13}$C and $^{14}$N through the CN cycle.
As a consequence, the $^{12}$C/$^{13}$C ratio decreases in the envelope to
values close to the CN-cycle equilibrium ratio ($\simeq$3$-$4). Theoretically,
HBB models also  predict the production of the short-lived $^{7}$Li through the
Cameron \& Fowler (1971) mechanism.

The chemistry of the dust in AGB circumstellar shells is also determined by the
C/O ratio. Some authors have considered a few carbon stars with circumstellar
amorphous silicate dust emission as transition objects between O-rich and
C-rich stars on the AGB (e.g. Chan \& Kwok 1991; Kwok \& Chan 1993). These
stars are generally classified as peculiar J-type stars, as  they show
characteristic low $^{12}$C/$^{13}$C ratios, the presence of Li enrichment and
no s-process element overabundances in their atmospheres (Abia \& Isern 2000),
but its evolutionary status is controversial. At present, they are thought to
be low-mass AGB stars (M $<$ 2 M$_\odot$) (Abia et al. 2003), but some AGB
stars with higher masses (M $\gtrsim$ 3$-$4 M$_\odot$) which are experiencing
HBB (e.g. Lorenz-Martins 1996) can show very similar properties.

Remarkably, the only known C-rich AGB star showing strong crystalline silicate
emission is IRAS 09425$-$6040 ( =GLMP 260; hereafter I09425). I09425 
displays the highest
proportion of crystalline silicates ($\sim$75 \%; i.e. only comparable to the
Hale-Bopp comet) observed in any source so far (Molster et al. 2001), while the
short wavelength part ($\lambda$ $\lesssim$ 15 $\mu$m) of the ISO spectrum is
dominated by deep absorption bands from C-rich gas-phase molecules (e.g.
C$_2$H$_2$, HCN, etc.). In this paper we present a chemical abundance analysis
of this star, based on high resolution spectra obtained during an optical
survey carried out on a large sample of massive galactic O-rich AGB stars
(Garc\'\i a-Hern\'andez et al. 2006). Section 2 describes the optical
observations performed and the results obtained. The main physical parameters
and the chemical abundances of I09425 are derived in Section 3 while its nature
and evolutionary stage is discussed in Section 4. 

\section{Optical Spectroscopic Observations}
The high resolution optical spectrum presented here was taken on 1997 February
23 with the CAsegrain Echelle SPECtrograph (CASPEC) of the ESO 3.6m telescope.
The integration time was 30 minutes and we used a TEK 1024 x 1024 CCD with a 24
$\mu$m pixel size. We used the 31 lines mm$^{-1}$ grating and continuously
covered the spectral range 6000-8300 $\AA$ at R$\sim$40,000 in about 27 orders,
with small gaps in the redder orders. The two-dimensional spectra were reduced
following the standard procedure for echelle spectroscopy using
IRAF\footnote{Image Reduction and Analysis Facility software is distributed by
the National Optical Astronomy Observatories, which is operated by the
Association of Universities for Research in Astronomy, Inc., under cooperative
agreement with the National Science Foundation.} astronomical routines. Note
that the few spectral ranges used in the abundance analysis presented in this
paper are not significantly affected by terrestrial features. The S/N ratio
achieved in the final spectrum varies from the blue to the red orders. At
$\sim$6000 $\AA$ the S/N ratio is 30$-$40, while at $\sim$8000 $\AA$ the S/N
ratio is higher than 100.

Based on the available low resolution spectrum (Su\'arez et al. 2006), the
source can be identified as an extremely reddened carbon star with H$\alpha$ in
emission. The steepness of the spectrum is such that no continuum is visible
shortwards 5500 $\AA$.

The high resolution optical spectrum of I09425 here discussed is dominated by
the presence of multiple CN and C$_{2}$ absorption bands. Overimposed, we can
still clearly identify some strong atomic lines such as the Li I resonance line
at 6708 $\AA$, the K I line at 7699 $\AA$, the Rb I line at 7800 $\AA$ and a 
few
atomic lines corresponding to s-process elements. Interestingly, we found that
the resonance atomic lines of Li I, Rb I and K I display blue-shifted
circumstellar components, which are stronger than the photospheric absorptions,
suggesting the presence of an expanding circumstellar envelope around I09425. 
We
derived Doppler velocities of $-$22.3, $-$21.2 and $-$21.8 km s$^{-1}$ from the
Li I, Rb I and K I circumstellar lines, respectively. In Figure 1 the optical
spectrum of I09425 from 6680 to 6720 $\AA$ is compared with the spectra of 
other
well studied C-rich AGB stars (the N-type star U Hya, the J-type star RY Dra 
and
the `super Li-rich' star IY Hya). The Li I line in I09425 is very strong, but
not as much as in IY Hya, indicating that possibly I09425 is not a `super
Li-rich' star. In Figure 1 we can also see the strong blend at 6687.6
\AA~corresponding to Y I which is very strong as well in IY Hya and in the
N-type carbon star (U Hya), but it is completely missed in the J-star RY Dra.
The s-process element atomic lines such as Y I at 6024, 6434 and 6793, Zr I at
5955, 6025, 6062, 6762, Ba I at 6142, La I at 6578 and 7334 and Gd I 7135, 
among
many others, are also very strong in the spectrum of I09425, suggesting that it
may be slightly enriched in s-elements (see Section 3.1).

\begin{figure}
\includegraphics[angle=-90,width=9cm]{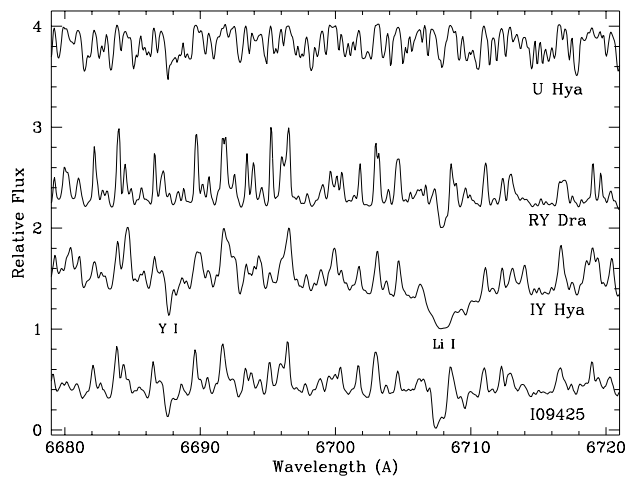}
\caption{Spectral region around the Li I line at 6708 $\AA$ of IRAS
09425$-$6040, compared with the same region in three other AGB carbon stars of
different type. The continuum level has been placed at the same point
($\sim$6725 \AA) in all the spectra. From top to bottom: the N-type star U Hya,
the J-type star RY Dra, the super Li-rich star IY Hya, and IRAS 09425$-$6040.
Note the strong similarity (except by the very different strength of the Li I
line) between the spectrum of IRAS 09425$-$6040 and that of IY Hya.}
\label{Fig.1}
\end{figure}

\section{Data Analysis}
\subsection{Chemical Abundances}
Very little photometric information is available in the literature about I09425
to estimate its atmospheric parameters. From the available infrared colours and
the calibration by Ohnaka \& Tsuji (1996), we estimate a very low
\textit{$T_{eff}$} = 2000 K which seems unrealistic. Synthetic spectra computed
with a C-rich model atmosphere with this effective temperature actually result
in very strong CN and C$_{2}$ absorption bands which are not seen in the
spectrum of I09425. This probably indicates that the photosphere of the star is
hotter and that infrared colours may be affected by the presence of host dust
in the circumstellar envelope. Thus, we computed synthetic spectra with
different values of the stellar parameters until a reasonable agreement and
consistency was found in the fit of several spectral ranges (6700 \AA~for Li,
7800 \AA~for Rb, and 8000 \AA~for the carbon isotopic ratio). The atmospheric
parameters finally adopted were \textit{$T_{eff}$} = 2850 K, \textit{log g} =
0.0, \textit{[Fe/H]} = 0.0 and \textit{$\xi$} = 2.2 km s$^{-1}$, which are
typical values for galactic carbon stars (e.g. Lambert et al. 1986). The
uncertainty in these parameters is high, the most important ones affecting the
errors in the abundance determination being \textit{$T_{eff}$} ($\pm$300 K) and
microturbulence ($\pm$1 km s$^{-1}$). The uncertainty in gravity ($\pm$0.5 dex)
and metallicity ($\pm$ 0.3 dex) do not significantly affect the derived
abundances. We have used the grid of model atmospheres for C stars computed by
the Uppsala group (see Eriksson et al. 1984 for details). The adopted atomic
line list is basically that used in Abia et al. (2001) after some revisions
using solar \textit{gf}-values derived by Th\'evenin (1990). The molecular line
list includes CN, C$_{2}$ and CH. The C$_{2}$ lines are from Querci, Querci \&
Kunde (1971) while that CN and CH lists were assembled from the best available
data and are described in Hill et al. (2002) and Cayrel et al. (2004).

By using the model atmosphere mentioned above and the synthesis spectral
technique, the abundance ratios  C/O = 1.01, $^{12}$C/$^{13}$C = 15$\pm$6 and
estimations of   the abundances of Li and Rb, namely: \textit{log (Li/H) +
12} $\sim$  0.7$\pm$0.4 and [Rb/Fe] $\sim$ 0.1$\pm$0.3 (which should be taken
cautiously,   because the lines used are affected by the presence of a
circumstellar  component) were derived\footnote{For details about the
technique used,  see Abia \& Isern (2000), Abia et al. (2001).}. Our
determination of the $^{12}$C/$^{13}$C ratio from the optical spectrum is in
good agreement with the $^{12}$C/$^{13}$C ratio of $\sim$10 derived by Molster
et al. (2001) at sub-millimetre wavelengths. The estimated errors reflect
mostly the sensitivity of the derived abundances to changes of the model
atmosphere parameters and do not consider possible non-LTE effects, dynamics of
the atmosphere, errors in the model atmosphere or errors in the
molecular/atomic linelists. The best fit around the spectral region used to
derive the Li abundance (6680-6730 \AA) is showed in Figure 2. From our
chemical analysis, we also found that the best fit in the spectral ranges
studied is always obtained with a slight overabundance of s-elements such as
Zr, Y, Ba, La, etc, and with a slightly metal-poor model atmosphere. However,
the large error bar associated to the derived s-element abundances prevent us
to reach any firm conclusion about the possible s-element overabundance until a
more detailed analysis is done.

\begin{figure}
\includegraphics[angle=-90,width=9cm]{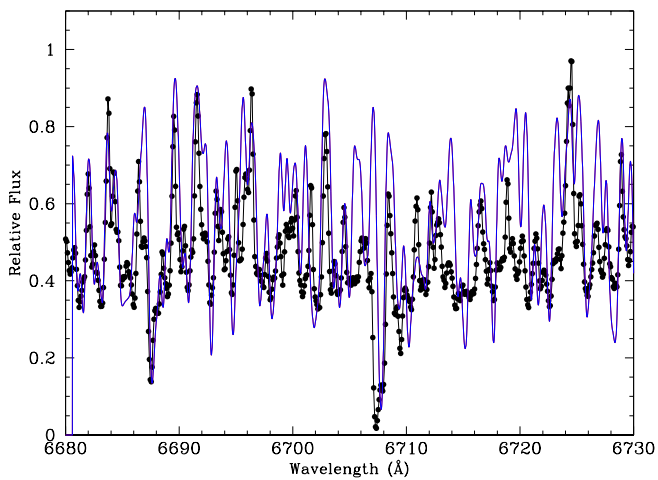}
\caption{Synthetic (blue) and observed (black) spectra of IRAS 09425$-$6040 in
the wavelength region around the Li I line at 6708 \AA. Note that the lithium
line has a clear ``blue-shifted'' circumstellar component.}
\label{Fig.2}
\end{figure}

\subsection{Other Data}
Unpublished optical HST/ACS images taken through the broad-band filters F435W
and F606W with several exposure times ranging from 1 to 250 s were also
retrieved from the HST Data Archive\footnote{They were originally part of the
snapshot program 9463 (P.I. Sahai)}. We found that I09425 is the brightest
source in the field in the F606W filter while it is barely detectable in the
F435W filter. This is consistent with the characterictics seen in the optical
spectrum of I09425. In addition, we found that I09425 is a point-like source to
HST (FWHM=0.09").

ISO SWS spectroscopy of I09425 has already been analysed in detail by Molster 
et
al. (2001) where its chemical dichotomy (C-rich and O-rich) became evident for
the first time. From the ground-based near-IR photometry taken during the 
period
1990-1992 and reported by Garc\'\i a-Lario et al. (1997) and Fouque et al.
(1992) complemented with more recent 2MASS and DENIS data taken in the period
1999-2000, we found a variability amplitude wider than 1.2, 0.9 and 0.7 mag in
the J, H and K bands, respectively. This large amplitude suggests that 
I09425 is
likely a long-period ($\sim$500$-$700 days) Mira-like star. On the other hand,
the red IRAS colours indicate that I09425 is a very red star which is probably
experiencing a strong mass loss. Consistently, Molster et al. (2001) derived a
mass-loss rate of 2x10$^{-6}$ M$_{\odot}$ yr$^{-1}$ from observations of the
pure rotational transitions of $^{12}$CO at J=1-0, J=2-1 and of $^{13}$CO at
J=2-1 at 115, 230 and 22 GHz, respectively. 

By integrating the observed flux at all wavelengths, a distance-dependent
luminosity of $\sim$1710 (D/kpc)$^{2}$ L$_\odot$ is obtained. I09425 is located
in the direction of the Carina star forming region which is at a distance of
$\sim$2.7 kpc (Tapia et al. 2003). If we assume that I09425 is also a member of
the Carina complex (and this is supported by the radial velocity derived from
our optical spectrum), a total luminosity of $\sim$12,500 L$_\odot$ is derived.
According to Bl\"ocker (1995) this would correspond to an AGB star with a core
mass of 0.84 M$_\odot$, suggesting a relatively massive progenitor (M $\gtrsim$
3$-$5 M$_\odot$). 

\section{Discussion}
Considering the above results, several are the possibilities to explain the
evolutionary status of I09425:

a) I09425 is a low-mass binary carbon star (M $<$ 2$-$3 M$_\odot$) and its
O-rich circumstellar shell is the result from previous accretion from a
companion star. As proposed by Molster et al. (2001), the O-rich dust could be
stored in a massive circumbinary disk. This would be consistent with the strong
crystalline silicate emission detected by ISO. However, the observed 12/2.2
$\mu$m flux ratio is more consistent with the star being an ``intrinsic'' (no
binary) AGB star (Jorissen et al. 1993) which has very  recently experienced a
very strong mass loss. Note that binary (extrinsic)  AGB S-type stars do not
usually show Li enhancements (e.g. Barbuy et al. 1992). It is unlikely that Li
could survive during the mass-transfer and subsequent mixing. In addition, no
firm evidence for binarity in I09425 nor for any disk structure exists yet. 

b) I09425 is a low-mass J-type carbon star. The peculiar chemical abundances of
I09425 are rather similar to the J-stars studied by Abia \& Isern (2000).
J-type stars are preferentially identified as low-mass AGB stars (M $<$ 2
M$_\odot$) on the early AGB (Abia et al. 2003). Some authors have suggested
that J-stars could bring up to the surface the C-rich material via a process
distinct to the third-dredge$-$up (e.g. the helium flash) (e.g. Deupree \&
Wallace 1996). In order to explain the Li and $^{13}$C production, a
non-standard mixing process, the so-called ``Cool Bottom Processing'' (CBP)
(Wasserburg, Boothroyd \& Sackmann 1995; Dom\'\i nguez et al. 2004) is invoked.
However, the strong infrared excess observed in I09425 together with the
properties of the Mira-like variability argue against an early AGB status for
this object. There are no other J-type stars known to exhibit crystalline
silicates (see e.g. Yang, Chen \& He 2004) which are only observed in high mass
loss rate O-rich AGB stars (see e.g. Sylvester et al. 1999).

c) I09425 is a TP-AGB star with a progenitor mass close to the limit for the
HBB occurrence ($\sim$3$-$4 M$_\odot$). This interpretation is consistent with
the observed Mira-like variability and with the strong IR excess observed,
which indicate that the star may be in an advanced stage on the AGB. This would
also be consistent with the estimated progenitor mass of $\sim$3$-$5 M$_\odot$,
previously derived from luminosity considerations. Under this scenario, I09425
would be a HBB AGB star which has lost an important fraction of its envelope
mass (O-rich) during its evolution, and in which HBB may have recently become
deactivated because of the strong mass loss (e.g. Frost et al. 1998) or simply
by the $^{3}$He exhaustion in the envelope (e.g. Forestini \& Charbonnel 1997)
but not the operation of the third dredge$-$up. Note that the probable slight
s-element enrichment, as suggested from our chemical analysis, is also in good
agreement with the small s-process element enhancement recently found in
massive (M $\gtrsim$ 3 M$_\odot$) galactic O-rich TP-AGB stars by Garc\'\i
a-Hern\'andez et al. (2006). After a few thermal pulses more, and according to
the theoretical models, the O-rich AGB star can turn into a C-rich star still
showing some Li and a low $^{12}$C/$^{13}$C ratio from the previous HBB epoch
(see Lattanzio \& Forestini 1999). For some time a mixed chemistry (a C-rich
central star and O-rich dust in the envelope) will be observed in its spectrum.

The crystallization of amorphous silicates occurs via the heating and the
subsequent cooling of the dust grains. This may occur slowly at low temperature
in a long-lived circumbinary disk, under the influence of UV radiation (as in
the Red Rectangle, although here the temperature of the central star is so low
that  the expected production of UV photons is negligible), or quickly in the
AGB wind at very high mass loss rates, through high-temperature annealing
(Waters et al. 1996; Sylvester et al. 1999). We suggest that this latter
mechanism may have operated in the case of I09425 to produce the huge
crystalline silicate emission that we now observe.

The observation of stars like I09425 is very improbable because this
evolutionary phase is short-lived and extremely rare. The detection of Li and
the presence of C-rich gas-phase molecules in the ISO spectrum of the central
star (and also the C/O ratio of 1.01) suggest that the change of chemistry has
been relatively recent because of Li has not been destroyed and there has been
no time to form more complex molecules such as PAHs. The bulk of the previously
expelled O-rich material would be now found only farther away from the central
star as cooler O-rich dust. Similar observational double-dust chemistry
properties are observed in a few transition sources from AGB stars and
planetary nebulae (the so-called post-AGB stars), which are interpreted as the
consequence of a late TP at the end of the AGB (e.g. Zijlstra 2001). It is
tempting to speculate that in I09425 we are observing the same physical process
at an earlier stage of evolution. Further observations in the near  future may
reveal  additional surprises and will certainly help clarifying the
evolutionary status and main properties of this  rather peculiar star, which
may be rapidly evolving towards the post-AGB stage.

\begin{acknowledgements}
AM and PGL acknowledge support from grant \emph{AYA 2004$-$3136 and
AYA 2003$-$9499 from the Spanish Ministerio de Educaci\'on y Ciencia}.
\end{acknowledgements}

{} 

\begin{thebibliography}{} 
\bibitem[2000]{Abia} Abia, C., \& Isern, J. 2000, ApJ, 536, 438
\bibitem[2001]{Abia} Abia, C., Busso, M., Gallino, R., Dom\'\i nguez, I., Straniero, O., \& Isern, J. 2001, ApJ, 559, 1117
\bibitem[2003]{Abia} Abia, C., Dom\'\i nguez, I., Gallino, R., Busso, M., Straniero, O ., de Laverny, P., \& Wallerstein, G. 2003, PASA, 20, 314
\bibitem[1992]{Barbuy92} Barbuy, B., Jorissen, A., Rossi, S. C. F., \& Arnould, M. 1992, A\&A, 262, 216
\bibitem[1995]{Blocker95} Bl\"ocker, T. 1995, A\&A, 299, 755 
\bibitem[1971]{Cam71} Cameron, A. G. W., \& Fowler, W.A. 1971, ApJ, 164, 111
\bibitem[2004]{Cayrel04} Cayrel, R., Depagne, E., Spite, M., Hill, V., Spite, F., François, P., Plez, B., Beers, T., Primas, F., Andersen, J., Barbuy, B., Bonifacio, P., Molaro, P., \& Nordstr$\ddot{o}$m, B. 2004, A\&A, 416, 1117
\bibitem[1991]{Chan} Chan, S. J., \&  Kwok, S. 1991, ApJ, 383, 837
\bibitem[1996]{Deupree96} Deupree, R.G., \& Wallace, R.J. 1996, ApJ, 317, 214
\bibitem[2004]{dominguez04} Dom\'\i nguez, I., Abia, C., Straniero, O., Cristallo, S., \& Pavlenko, Ya. V. 2004, A\&A, 422, 1045
\bibitem[1984]{erik84} Eriksson, K., Gustafsson, B., J\"{o}rgensen, U. G., \& Nordlund, A. 1984, A\&A, 132, 37
\bibitem[1997]{For97} Forestini, M., \& Charbonnel, C. 1997, A\&AS, 123, 241
\bibitem[1992]{Fouque} Fouque, P., Le Bertre, T., Epchtein, N., Guglielmo, F., \& Kerschbaum, F. 1992, A\&ASS, 93, 151
\bibitem[1998]{Frost98} Frost, C. A., Cannon, R. C., Lattanzio, J. C., Wood, P. R., \& Forestini, M. 1998, A\&A, 332, L17
\bibitem[2006]{gh06} Garc\'\i a-Hern\'andez, D. A., Garc\'\i a-Lario, P., Plez, B., Manchado, A., D'Antona, F., Lub, J., \& Habing, H. 2006, A\&A (submitted)
\bibitem[1997]{gl97} Garc\'\i a-Lario, P., Manchado, A., Pych, W., \& Pottasch, S. R. 1997, A\&ASS, 126, 479
\bibitem[2002]{hill02} Hill, V., Plez, B., Cayrel, R., Beers, T. C., Nordstr$\ddot{o}$m, B., Andersen, J., Spite, M., Spite, F., Barbuy, B., Bonifacio, P., Depagne, E., François, P., \& Primas, F. 2002, 387, 560
\bibitem[1993]{jorissen93} Jorissen, A., Frayer, D. T., Johnson, H. R., Mayor, M., \& Smith, V. V. 1993, A\&A, 271, 463
\bibitem[1993]{Kwok} Kwok, S., \& Chan, S. J. 1993, AJ, 106, 2140
\bibitem[1986]{Lambert86}Lambert, D. L., Gustafsson, B., Eriksson, K., \& Hinkle, K. H. 1986, ApJS, 62, 373
\bibitem[1999]{Latt99} Lattanzio, J.L. \& Forestini, M. 1999, in ``{\it AGB Stars, IAU Symposium No. 191}'', Eds. T. Le Bertre, A. L\'ebre and C. Waelkens, p. 31
\bibitem[1996]{Lorenz} Lorenz-Martins, S. 1996, A\&A, 314, 209
\bibitem[1999]{Mazzitelli} Mazzitelli, I., D'Antona, F., Ventura, P. 1999, A\&A, 348, 846
\bibitem[2001]{Molster01} Molster, F. J., Yamamura, I., Waters, L. B. F., Nyman, L. -$\AA$., K$\ddot{a}$ufl, H. -U., de Jong, T., \& Loup, C. 2001, A\&A, 366, 923
\bibitem[1996]{Ohna96} Ohnaka, K., \& Tsuji, T. 1996, A\&A, 310, 933
\bibitem[1971]{Querci71} Querci, F., Querci, M., \& Kunde, V. G. 1971, A\&A, 15, 256
\bibitem[1992]{Sackmann92} Sackmann, I.-J. \& Boothroyd, A.I. 1992, ApJ, 392, L71
\bibitem[2005]{sua05} Su\'arez, O., Garc\'{\i}a-Lario, P., Manchado, A., Manteiga, M., Ulla, A, \& Pottasch, S. R. 2006, A\&A (submitted)
\bibitem[2003]{Tapia03} Tapia, M., Roth, M., V\'azquez, Rub\'en A., \& Feinstein, A. 2003, MNRAS, 339, 44
\bibitem[1999]{Sylvester99} Sylvester, R. J., Kemper, F., Barlow, M. J., de Jong, T., Waters, L. B. F. M., Tielens, A. G. G. M., \& Omont, A. 1999, A\&A, 352, 587
\bibitem[1990]{Thev90} Th\'evenin, F. 1990, A\&AS, 89, 179
\bibitem[1995]{Wasserburg95} Wasserburg, G. J., Boothroyd, A. I., \& Sackmann, I. J. 1995, ApJ, 447, L37
\bibitem[1996] {Waters96} Waters, L.B.F.M., Molster, F.J, de Jong,
T. et al. 1996, A\&A 315, L361 
\bibitem[2004]{Yang04} Yang, X., Chen, P., \& He, J. 2004, A\&A, 414, 1049
\bibitem[2001]{Zijlstra01} Zijlstra, A. A. 2001, Ap\&SS, 275, 79

\end{thebibliography}
\end{document}